\documentstyle[12pt]{article}
\def\kph{{KP hierarchy}}
\def\kdvh{{KdV hierarchy}}
\def\a{\begin{eqnarray}}
\def\b{\end{eqnarray}}
\def\ba{\begin{array}}
\def\ea{\end{array}}
\def\0{\nonumber}
\def\d{{\partial}}

\def\dt{{\partial \over {\partial t}}}
\def\ddtr{{\partial \over {\partial t_r}}}

\begin{document}
\begin{flushright}
SISSA-ISAS 57/93/EP\\
ASITP 93--25\\
hep-th/9305005
\end{flushright}
\vskip0.5cm
\centerline{\LARGE\bf Multi--field representations of the \kph}
\vskip0.5cm
\centerline{\LARGE\bf  and multi--matrix models}
\vskip2.3cm
\centerline{\large  L.Bonora}
\centerline{International School for Advanced Studies (SISSA/ISAS)}
\centerline{Via Beirut 2, 34014 Trieste, Italy}
\centerline{INFN, Sezione di Trieste.  }
\vskip0.5cm
\centerline{\large C.S.Xiong}
\centerline{Institute of Theoretical Physics, Academia Sinica}
\centerline{P.O.Box 2735, Beijing 100080, China}
\vskip6cm
\abstract
{We discuss the integrable hierarchies that appear in multi--matrix models.
They
can be envisaged as multi--field representations of the \kph. We then study
the possible reductions of this systems via the Dirac reduction method
by suppressing successively one by one part of the fields.
We find in this way new integrable hierarchies, of which we are able to
write  the Lax pair representations by means of suitable
Drinfeld--Sokolov linear systems. At the bottom of each reduction
procedure we find an $N$--th \kdvh. We discuss in detail the case which
leads to the KdV hierarchy and to the Boussinesque hierarchy,
as well as the general case in the dispersionless limit.}
\vfill
\eject

\section{Introduction}

In \cite{BX2} we showed that in multi--matrix models there naturally appear
pseudo--differential operators of the type (up to some relabelling)
\a
L=\d +\sum_{l=1}^{N-1} a_l \frac{1}{\d -S_l}
\frac{1}{\d -S_{l-1}}\cdots \frac{1}{\d -S_1}\label{defL}
\b
where $\d=\frac{\partial}{\partial x}=\frac{\partial}{\partial t_{1,1}}$
and $a_1,..., a_{N-1}, S_1,...,
S_{N-1}$
are independent
coordinates ({\it fields}) of the system (throughout the paper, when we write
$N$, we refer precisely to this equation).
They appear in linear systems of the type
\a
&&L \Psi =\lambda \Psi\label{LS}\\
&&\ddtr \Psi = (L^r )_+ \Psi, \quad\quad r=1,2,3,....\label{flows}
\b
whose consistency conditions are
\a
\ddtr L= [ (L^r)_+, L] \label{CC}
\b
Here $\lambda$ is the spectral parameter and the $t_r\equiv t_{1,r}$'s
are, for example, the self--coupling parameters of the
first matrix in the model.
In the two above equations, the label + appended to a pseudo--
differential operator represents the purely differential part of it.

In \cite{BX2} we proved that multi--matrix models are {\it exactly} represented
by the solutions of the above integrable hierarchies of equations that moreover
satisfy suitable string equations (we do not write them down here since they
do not enter the subsequent discussion). This sets the problem of
classifying the various distinct integrable hierarchies that come up in
this way and, subsequently, the problem of identifying these hierarchies
with topological models of 2d gravity plus matter, just as we did in
\cite{BX1} for one--matrix models. Here we will be dealing only with the
first problem: precisely we want not only to describe the integrable
hierarchies
of the type (\ref{CC}), but also those that can be obtained
from them via reduction.

The main result of this paper can be summarized as follows: 1) for each $N$ in
(\ref{defL}) we find $N$ distinct integrable hierarchies which are obtained
by suppressing successively the fields $S_l$; 2) of each such hierarchy
we write down a Drinfeld--Sokolov linear system from which the Lax pair
representation can be easily extracted; 3) it is irrelevant what fields
we suppress
first, the hierarchies with the same number of $S$ fields being
isomorphic; 4) at the end of this cascade
procedure we find the $N$--th \kdvh. This is likely to represent a complete
classification of all the integrable hierarchies (with a finite number of
fields) appearing in multi--matrix models.

As for the reduction procedure, it
is not simply a restriction on the flow equations (as we did in \cite{BX1}
for the NLS hierarchy), but a true reduction of the Hamiltonian system
where the vanishing of the $S$ fields is imposed as a second class
constraint and leads to the introduction of the corresponding Dirac brackets
\footnote{Thoughout the paper we keep this technical distinction: restriction
means imposing some constraints on the equations of motion, while reduction
means Hamiltonian reduction.}.

We remark that this result is, in particular, a proof of the conjecture
that multi--matrix models do lead to $N$--th KdV hierarchies, \cite{D}.
One should also add that the result is interesting independently
of multi--matrix
models: in fact the hierarchy (\ref{LS},\ref{CC}) is an integrable restriction
of the \kph\, which goes under the name of multi--field representation of
the \kph. The properties of the multi--field representations of the
\kph\, , not to speak of their reductions, have not received thus
far a large attention, see however \cite{K},\cite{OS},\cite{ANPV},\cite{BIK}.
But, needless to say, for us the main motivation to study these problems
came from multi--matrix models.

The letter is organized as follows. In section 2 we rediscuss the reduction
from
the NLS hierarchy to the KdV hierarchy as a prototype of the subsequent
reductions.
In section 3 we discuss at length the system corresponding to $N=3$ and in
section 4 its integrable reduction. In section 5 we present a compact way
to describe all the integrable systems of the type (\ref{defL})  together
with their integrable reductions. In section 6 we discuss the dispersionless
limit of the system (\ref{defL}) with generic $N$ and its integrable
reduction, which turn out to be
the dispersionless limit of the $N$--th \kdvh.

To end this section let us introduce a few definitions. Throughout the
paper we come across integrable systems which can be given
a Lax pair representation, i.e. they can be formulated by means of
a differential or pseudo--differential
operator of the form
\[
A= \d^k+ a\d^{k-2}+b\d^{k-3}+...
\]
The corresponding flows then are given by
\a
\ddtr A=[A_+^{\frac{r}{k}}, A]\label{flow}
\b
We define
\a
<A>=\int\/dx a_{-1}(x),\qquad\qquad
A=\ldots+a_{-1}(x)\partial^{-1}+\ldots\0
\b
and for any differential
operator $X$ we set $f_X(A)= <AX>$. Two Poisson brackets
are then given by
\a
\{f_X, f_Y\}_1(A)&=&<A[Y, X]_R>\label{LPB1}\\
\{f_X, f_Y\}_2(A)&=&<(XA)_+YA>
-<(AX)_+AY>+\label{LPB2}\\
&&+\frac{1}{k} \int [A,Y]_{(-1)}\Big(\d^{-1}[A,X]_{(-1)}\Big)\0
\b
In the first bracket, \cite{STS}, \cite{Dik},
\[
\relax [X,Y]_R= \frac {1}{2} \Big( [RX,Y]+[X,RY]\Big),\quad\quad\quad
RX= X_+-X_-, \quad\quad X_-=X-X_+
\]
and in the second bracket
the label $\textstyle (-1)$ means the coefficient of $\d^{-1}$.
We can further introduce the Hamiltonians via
\a
H_r={k\over r}<A^{r\over k}>, \qquad\qquad \forall k\geq1\0
\b
Integrability of the system is guaranteed since these quantities are in
involution and the Poisson brackets are compatible, i.e.
\[
\{H_{r+k},f\}_1=\{H_r, f\}_2\qquad\hbox{\rm for any function}\quad f
\]
In particular the flow equations are given by
\a
\ddtr f= \{f,H_r\}_2= \{f,H_{r+k}\}_1\label{hflow}
\b

\section{Reduction of the NLS hierarchy}

To illustrate our method we rederive here the reduction of the NLS
hierarchy to the \kdvh. One Hermitean matrix models with the most
general potential are exactly described by the NLS hierarchy
\cite{BX},\cite{BX1}.

The NLS hierarchy can be defined via the Lax operator
\footnote{We remark that this differs from the Lax operator used in \cite{BX1}
by the redefinition $S \rightarrow S+ (\ln R)'$. We adopt the definition
(\ref{NLS}) in order to conform to the general definition (\ref{defL})
-- this is the case $N=2$ with $a_1=R$ and $S_1=S$.}
\a
L=\d +R\frac{1}{\d-S}\label{NLS}
\b
Now,
following the prescription of the previous section, we can calculate
anything we need. In particular the first Poisson brackets are
\a
&&\{R(x), R(y)\}_1=0\qquad\{S(x), S(y)\}_1=0\0\\
&&\{R(x), S(y)\}_1=-\partial_x\delta(x-y)\0
\b
and the second ones are
\a
&&\{R(x), R(y)\}_2=\bigl(2R\partial+R^{'}\bigl)\delta(x-y)
\label{RR2}\\
&&\{S(x), S(y)\}_2=2\partial\delta(x-y)\label{SS2}\\
&&\{R(x), S(y)\}_2=\bigl(S\partial+\partial^2\bigl)\delta(x-y)
\label{RS2}
\b
Henceforth we will use for Poisson brackets the simplified notation
of (\ref{RS2}) where the fields and derivatives (denoted also by a prime)
appearing in RHS are understood to be evaluated at $x$.

The Hamiltonians are
\[
H_1=\int dx R,\quad\quad
H_2=\int dx RS,\quad\quad
H_3=\int dx (R^2+RS^2+R'S), \quad{\rm etc.}
\]
and the flow equations are easy to compute. For example
the $t_2$ flow equations are:
\a
\frac {\partial S}{\partial t_2}= -S'' +2S'S + 2R',\quad\quad
\frac {\partial R}{\partial t_2}= -3R'' +2(RS)' \label{2flow}
\b
which is a disguised form of the non--linear
Schr$\ddot {\rm o}$dinger equation.

Let us study now the reduction
\[
S=0
\]
In \cite{BX1} we considered a restriction $S=0$ applied only to the flow
equations and we ended up with the KdV hierarchy. However the
restriction method, tough valid in this case, leads to inconsistent
results when $N>2$ in (\ref{defL}), \cite{BX2}. We have to
consider the reduction of the Hamiltonian system. The constraint $S=0$
is second class and we have to follow Dirac's procedure, \cite{Dirac}. For the
second Poisson brackets we can introduce the corresponding Dirac
bracket in the reduced system
\a
\{R(x), R(y)\}_{2D}=\bigl(2R\partial+R^{'}
+\frac{1}{2} \d^3\bigl)\delta(x-y)
\label{RRD}
\b
If we use this and the constrained $H_3$ Hamiltonian we obtain
\a
{\partial\over{\partial t_3}}R=R^{'''}
+6RR^{'}\label{KdV}
\b
which is the KdV equation.

The first Poisson bracket is not reducible. So we have to look for
another Poisson bracket for the reduced system. It is not
difficult to find it as well as the series of Hamiltonians using the fact
that the two Poisson brackets must be compatible and the
Hamiltonians in involution. This is however a long procedure.
There is a much quicker method, which consists in finding
a Lax pair representation. For this particular case the solution is
universally known: the Lax operator is (up to a rescaling of the flow
parameters)
\[
\d^2+R
\]
With this and the formulas of the introduction we can calculate
Poisson brackets, Hamiltonians and flows (i.e. the
KdV hierarchy) of the reduced system.

This suggests us a general method to study the reductions
of the more complex systems with $N>2$. We will impose
restriction constraints which will turn out to be second class.
We will compute the Dirac brackets corresponding to the
second Poisson structure and find the equations of
motion of the reduced system. It will be then easy to identify the appropriate
Lax operator.

\section{Four--field representation of the \kph}

The simplest integrable system that appear in matrix models
after the NLS system is the four--field restriction of the \kph. It
naturally leads, via reduction, to the Boussinesque hierarchy.
Let us describe this system in some detail.

We begin with the following set of differential equations
\a
&&\dt a_1=a_1^{''}+2a_2'+2(a_1 S_1)'\0\\
&&\dt a_2=a_2^{''}+2a'_2S_2+2a_2(S_1+S_2)'\label{3.1}\\
&&\dt S_1=2a'_1+2S_1S_1'-S_1^{''}\0\\
&&\dt S_2=2a'_1+2 S_2 S_2'-S_2^{''}-2 S_1^{''}\0
\b
We want to show that these equations do define an integrable
system, i.e. they possess a bi--Hamiltonian structure.

\subsection{The Poisson Brackets}

It is not difficult to show that these equations can be written as
Hamiltonian equations
\a
\dt f=\{f, H\}_1
\b
where the function $f$ stands for ($a_1, a_2, S_1, S_2$), and the Hamiltonian
takes the following form
\a
H=\int dx[a_1 S_1^2-a_1S'_1+a_2(S_1+S_2)+a_1^2]\label{h3}
\b
The Poisson brackets are
\a
&&\{a_1(x), S_1(y)\}_1=\delta'(x-y),\qquad
\{a_1(x), S_2(y)\}_1=\delta'(x-y),\0\\
&&\{a_2(x), a_2(y)\}_1=(2a_2\d+a'_2)\delta(x-y),\label{PB1}\\
&&\{a_2(x), S_2(y)\}_1=\Bigl(\d^2+(S_2-S_1)\d\Bigl)\delta(x-y).\0
\b
All the others vanish. One can check that these Poisson brackets satisfy
the Jacobi identity.

The important thing is that eq.(\ref{3.1}) admits another Hamiltonian
representation, that is
\def\th{{\tilde H}}
\a
\dt f=\{f, \th\}_2
\b
with
\a
\th=\int dx[a_1S_1+a_2]\label{h2}
\b
And the Poisson brackets take the form
\a
&&\{a_1(x), a_1(y)\}_2=(2a_1\d+a'_1)\delta(x-y),
\quad\quad\{a_1(x), a_2(y)\}_2=(3a_2\d+2a'_2)\delta(x-y),\0\\
&&\{a_1(x), S_1(y)\}_2=(\d^2+S_1\d)\delta(x-y),
\quad\quad\{a_1(x), S_2(y)\}_2=(2\d^2+S_2\d)\delta(x-y),\0\\
&&\{a_2(x), a_2(y)\}_2=[(2a'_2+4a_2S_2-2a_2S_1)\d
+a^{''}_2+(2a_2S_2-a_2S_1)']
\delta(x-y),\label{PB2}\\
&&\{a_2(x), S_2(y)\}_2=\Bigl(a_1\d+(\d+S_2)(\d+S_2-S_1)\d\Bigl)
\delta(x-y),\0\\
&&\{S_1(x), S_1(y)\}_2=2\delta'(x-y),\quad\quad\quad\{a_2(x), S_1(y)\}_2=0,\0\\
&&\{S_1(x), S_2(y)\}_2=\delta'(x-y), \quad\quad\quad\{S_2(x), S_2(y)\}_2
=2\delta'(x-y),\0
\b
It is straightforward to check that the Jacobi identities are satisfied.
Moreover
\[
\{H,\tilde H\}=0
\]
for both brackets. Therefore we have shown that the system (\ref{3.1}) has
a bi--Hamiltonian structure.

\subsection{Lax pair representation}

In principle we can use the two compatible Poisson brackets to construct
all the conserved quantities, the first few are
\a
H_1=\int dx a_1,\qquad H_2=\th,\qquad, H_3=H,\ldots\0
\b
In turn these conserved quantities generate new flows. The first ones
are trivial, the second flows are eq.(\ref{3.1}), and the third ones
take the following form
\a
&&\frac{\partial}{\partial t_3} a_1=[3a'_2+a^{''}_1+3a^2_1+3a'_1S_1+3a_1S_1^2
+3a_2(S_1+S_2)]'\0\\
&&\frac{\partial}{\partial t_3}  a_2=[a^{''}_2+3a'_2S_2+3a_2S_2^2]'
+6a'_1a_2+3a_1a'_2
+3a_2(S_1+S_2)S_1'-3a_2 S^{''}_1\0\\
&&\frac{\partial}{\partial t_3}  S_1=[3a_2+6a_1S_1+S_1^{''}+S^3_1-3S_1S'_1]'\\
&&\frac{\partial}{\partial t_3}  S_2=[3a_2-3a_1'+3a_1(S_1+S_2)+S_2^{''}
+3S_1^{''} +S_2^3-3S_2S_2'-3(S_1+S_2)S_1']'\0
\b
However there is a more way to proceed. It is by means of
the Lax pair representation. Consider
\a
L=\d+a_1{1\over{\d-S_1}}+a_2{1\over{\d-S_2}}{1\over{\d-S_1}}.
\label{lax4}
\b
Then not only can the above equations of motion be recast into
\a
\ddtr L=[L^r_+, L], \qquad r\ge 1.\label{kph4}
\b
with
the second flow ($r=2$) giving exactly
eq.(\ref{3.1}) with $t=t_2$ and the third flow ($r=3$)
giving the above equation with $t=t_3$, but, more important, we can compute
all the Poisson brackets, Hamiltonians and flows by means of the formulas
given in the introduction. The operator (\ref{lax4}) is nothing but our initial
operator (\ref{defL}) taken from multi--matrix models when $N=3$.

Comparing with the ordinary coordinates of \kph
\a
L=\d+\sum_{l=1}^{\infty}u_l\d^{-l}\label{laxkp}
\b
we see that our integrable system is nothing but a particular restriction
of the \kph, the restriction conditions being
\a
&&u_1=a_1, \qquad u_2=a_2+a_1 S_1\0\\
&&u_{l+2}=a_1(-\d+S_1)^{l+1}\cdot1+a_2\sum_{k=0}^l(-\d+S_2)^k
(-\d+S_1)^{l-k}
\cdot1.\quad l\geq 0 \label{kpcoor}
\b
So we may say that our system (\ref{kph4}) is nothing but the four field
representation of the \kph.

\subsection{$W_{\infty}$--algebra}

Since the new hierarchy is a restricted \kph\, we would
expect that the KP coordinates $u_l$'s defined in eq.(\ref{kpcoor}) to
form a $W_\infty$ algebra with respect to two Poisson brackets (\ref{PB1})
and (\ref{PB2}) (see in this regard, \cite{AFGZ},\cite{YW},\cite{P}).
This is actually the case. We give here only the result
with respect to the first Poisson bracket.

With respect to the coordinates $u_l$ of eq.(\ref{kpcoor})
\a
\{u_{i+2}(x), u_{j+2}(y)\}_1
=\bigg[\sum_{l=0}^{j+1}
\left(\ba{c} j+1\\ l\ea\right)\d^l u_{i+j-l+3}
-\sum_{l=0}^{i+1}(-1)^l
\left(\ba{c} i+1\\ l\ea\right)u_{i+j-l+3}\d^l\bigg]\delta(x-y)\0
\b
Therefore the $u_i$'s form a $W_{1+\infty}$-algebra, which we call
the four--field representation of such an algebra.

\section{Reductions of the four-field \kph}

\subsection{First reduction}

Let us consider the possible reductions of the integrable system
introduced in the previous section.
We will first impose the reduction
\a
S_1=0, \label{1rest}
\b
With respect to the second Poisson structure (\ref{PB2})
this is a second class constraint. However $\{S_1(x),S_1(y)\}$ has an inverse.
We can therefore proceed according to
Dirac and define improved Poisson brackets. These are
\a
&&\{a_1(x), a_1(y)\}_{2D}=(2a_1\d+a'_1+\frac{1}{2}\d^3)\delta(x-y),\0\\
&&\{a_1(x), a_2(y)\}_{2D}=(3a_2\d+2a'_2)\delta(x-y),\0\\
&&\{a_1(x), S_2(y)\}_{2D}=(\frac{3}{2}\d^2+S_2\d)\delta(x-y),\0\\
&&\{a_2(x), a_2(y)\}_{2D}=[(2a'_2+4a_2S_2)\d
+a^{''}_2+2(a_2S_2)']
\delta(x-y),\0\\
&&\{a_2(x), S_2(y)\}_{2D}=\Bigl(a_1\d+(\d+S_2)^2\d\Bigl)
\delta(x-y),\0\\
&&\{S_2(x), S_2(y)\}_{2D}=\frac{3}{2}\delta'(x-y)\label{PB2r1}
\b
Now the reduced Hamiltonian is a restriction of
the original Hamiltonian (\ref{h2}) to the constrained manifold.
Consequently the equations of motion of the constrained system are
\a
&&\dt a_1=2a_2'\0\\
&&\dt a_2=a_2^{''}+2(a_2S_2)'\label{3.1r}\\
&&\dt S_2=a'_1+2 S_2 S_2'-S_2^{''}\0
\b

The first Poisson bracket (\ref{PB1}) is not reducible.
One can then proceed to search for another Poisson structure, as we did
at the beginning of the previous section. One finds that with
the following choice
\a
\{a_1(x), a_1(y)\}_{1'}=2\d\delta(x-y),\qquad\qquad
\{a_2(x), S_2(y)\}_{1'}=\d\delta(x-y).\label{PB1r}
\b
and with the Hamiltonian
\a
H'= \int dx\Big( a_2(S_2^2-S_2')+a_1a_2\Big)\label{H'}
\b
one obtains the same equations of motion (\ref{3.1r}) and the two
Hamiltonians commute.
However,  as we have repeatedly noticed,
there is a much quicker and powerful way to
see the integrability of the reduced system. In fact we can write
down a Lax pair for it. The latter is based on the
following pseudo-differential operator
\a
\tilde L= \d^2 + a_1 +a_2 \frac{1}{\d-S_2}\label{tildeL}
\b
By means of the formulas in the introduction we can compute
the corresponding Poisson brackets, Hamiltonians and flow equations.
Among the latter we find all the above results. We conclude that the
reduced system is integrable \footnote{Via the transformation $\d \rightarrow
\d+S_2$ we obtain from (\ref{tildeL}) an operator which was considered
in \cite{K}}.

Next we consider another possible reduction. Instead of $S_1=0$ we impose
\[
S_2=0
\]
We proceed in the same way as above and we quickly realize that
the reduced system we obtain is nothing but a redefinition of the
first reduced system just defined. In fact if, starting from it, we make the
field redefinitions
\[
S_2\rightarrow S_1,\qquad\qquad a_1\rightarrow a_1 + S_1',
\quad\quad a_2 \rightarrow a_2+a_1S_1+S_1''+S_1S_1'
\]
we find the $S_2=0$ reduced system, which, consequently, is the same
as the first one.

Next one could wonder whether another reduction is possible, say
\[
S_1=S_2
\]
This reduction can be carried out, the Dirac bracket defined and
the equations of motion derived. However there does not seem to exist
any bi--Hamiltonian structure for this system. We will
henceforth ignore it.

\subsection{Further reduction. The Boussinesque hierarchy.}

In the previous subsection we have found one distinct integrable reduction
of the four--field \kph\, to a three field hierarchy. Here we further
reduce the reduced hierarchy by suppressing the remaining $S$ field
\a
S_2=0\label{2rest}
\b
With Dirac's procedure we find from (\ref{PB2r1})
\a
&&\{a_1(x), a_1(y)\}_{2D'}=(2a_1\d+a'_1+2\d^3)\delta(x-y),\0\\
&&\{a_1(x), a_2(y)\}_{2D'}=\Big(3a_2\d+2a'_2-\d^2 a_1+\d^4\Big)\delta(x-y),\0\\
&&\{a_2(x), a_2(y)\}_{2D'}=[2a'_2\d+a^{''}_2-\frac{2}{3}(a_1+\d^2)(\d
a_1+\d^3)]
\delta(x-y),\label{PB2r2}
\b
This is nothing but the $W_3$ algebra.
The corresponding equations of motion calculated from the doubly constrained
Hamiltonian (\ref{h2}) are
\a
\dt a_1=2a_2'-a_1'',\quad\quad\quad
\dt a_2=a_2^{''}-\frac{2}{3}(a_1a_1'+a_1'')\label{Bouss}
\b
This is known as the Boussinesque equation and it is the first of an integrable
hierarchy of equations which can be given a Lax pair representation
by means of the operator
\a
\tilde{\tilde L}= \d^3 + a_1 \d +a_2\label{Boussop}
\b
We remark that starting from the four--field representation of the \kph
and performing the simultaneous reduction $S_1=0=S_2$ we end up with the
same system (\ref{Boussop}).

In conclusion  as a result of the successive reduction of the four--field
\kph\, we find two more distinct integrable hierarchy. The last one
is the well--known Boussinesque hierarchy.

\section{A DS representation of the previous systems.}

The integrable models and reductions we have been considering so far
can be synthesized in a very compact and useful form that lends itself
to generalization via suitable Drinfeld--Sokolov (DS) linear systems. From
them one can easily extract the corresponding Lax pair.
Let us start with the linear system
\a
\left(\matrix{\d -S & -1\cr R &\d -\lambda\cr}\right)
\left(\matrix{ \psi_1\cr \psi\cr}\right)=0\label{LS21}
\b
{}From this it is elementary to see
\a
\big(\d+R\frac{1}{\d -S}\big)\psi=\lambda \psi\label{SE21}
\b
from which we recognize the spectral equation for the NLS Lax operator
studied in section 2.

Similarly
\a
\left(\matrix{\d  & -1\cr R-\lambda &\d \cr}\right)
\left(\matrix{ \psi\cr \psi_1\cr}\right)=0, \quad\quad {\rm and} \quad\quad
\Big(\d^2+R\Big)\psi= \lambda \psi\label{LS22}
\b
The last is nothing but the spectral equation for the KdV Lax operator.

Let us pass now to the systems with $N=3$ studied in section 3.
We have
\a
\left(\matrix{\d -S_2 & -1&0\cr 0&\d-S_1&-1\cr a_2&a_1 &\d-\lambda \cr}\right)
\left(\matrix{ \psi_1\cr \psi_2\cr\psi\cr}\right)=0\label{LS31}
\b
so that
\a
\Big(\d +a_1 \frac{1}{\d-S_1} +a_2 \frac{1}{\d-S_2}\frac{1}{\d-S_1}\Big)
\psi=\lambda\psi\label{SE31}
\b
Next
\a
\left(\matrix{\d -S_2 & -1&0\cr 0&\d&-1\cr a_2&a_1-\lambda &\d \cr}\right)
\left(\matrix{ \psi_1\cr \psi\cr\psi_2\cr}\right)=0,\quad\quad {\rm
and}\quad\quad
\Big(\d^2 +a_1 +a_2 \frac{1}{\d-S_2}\Big)\psi=\lambda\psi\label{SE32}
\b
Finally
\a
\left(\matrix{\d  & -1&0\cr 0&\d&-1\cr a_2-\lambda&a_1 &\d\cr}\right)
\left(\matrix{ \psi\cr \psi_1\cr\psi_2\cr}\right)=0,\quad\quad {\rm and}
\quad\quad
\Big(\d^3 +a_1 \d +a_2 \Big)\psi=\lambda\psi\label{SE33}
\b

We recognize the Lax operators $L,\tilde L$ and $\tilde {\tilde L}$ of
section 3. We notice that of all the above DS linear systems, only (\ref{LS22})
and (\ref{SE33}) are in the standard form \cite{DS}.

On the basis of these examples it is easy to extend this
to a generic $N$. The initial system is
\a
\left(\matrix{\d -S_{N-1} & -1&0&\cdots&0&0\cr
0&\d-S_{N-2}&-1&\cdots&0&0\cr
\cdots&\cdots&\cdots&\cdots&\cdots&\cdots\cr
0&0&0&\cdots&\d-S_1&-1\cr
a_{N-1}&a_{N-2}&a_{N-3}&\cdots&a_1 &\d-\lambda \cr}\right)
\left(\matrix{ \psi_1\cr \psi_2\cr\cdots\cr\psi_{N-1}\cr\psi\cr}\right)=0
\label{LSN1}
\b
{}From this we obtain the equation $L\psi=\lambda \psi$ where $L$ is the same
Lax operator as in eq.(\ref{defL}). The other systems can be obtained
by: 1) moving $-\lambda$ by one place to the left in the last row, 2)
suppressing the surviving S field with lowest index, 3) making an elementary
cyclic rotation upward of the vector entries, and 4) extracting the equation
for $\psi$.

We obtain in such a way $N-1$ more Lax operators. The obvious conjecture is
that these represent the $N-1$ distinct integrable reductions of the system
defined by (\ref{defL}). We uniquely identify each such system with a symbol
${\cal S}_N^k$, where $N$ has the usual meaning, eq.(\ref{defL}), and
$k$ counts the number of nonvanishing $S$ fields, $0\leq k\leq N-1$. In
particular the case $k=N-1$ yields the $2(N-1)$--field representation
of the KP hierarchy and of the $W_\infty$ algebra.

The above conjecture is very plausible, but we defer the
technicalities of the proof to a future publication.
Instead, in this letter, we study,
as additional evidence for our conjecture, the dispersionless limit of the
system ${\cal S}_N^{N-1}$
and its reduction. This limit is interesting in itself, firstly because
it represent a
new integrable hierarchy and secondly because, in the context of matrix
models, it represents the spheric limit of the theory.

\section{The dispersionless limit of ${\cal S}_N^{N-1}$}

The second flow equations of the Lax operator $L$ defined by eq.(\ref{defL}),
\a
\dt L=[L^2_+, L],\label{kphn}
\b
involving the fields $a_l, S_l; 1\leq l\leq N-1$, are
\a
&&\dt a_l=a^{''}_l+2a'_{l+1}+2a'_lS_l+2a_l(\sum_{k=1}^l S_k)',\0\\
&&\dt S_l=2a'_1+2S_lS'_l-S^{''}_l-2(\sum_{k=1}^{l-1} S_k)^{''}.\label{3.1gen}
\b

The recipe to define the dispersionless version of this integrable differential
system is very simple. We discard all
the higher derivatives in the equations or, alternatively, we make the
substitution
\a
&&\d\Longrightarrow p:{\rm canonical~ momentum},\label{presc2}\\
&&[\d, x]=1\Longrightarrow \{p, x\}=1.\0
\b

Applying the first procedure to eq.(\ref{3.1gen}), we obtain
\a
&&\dt a_l=2a'_{l+1}+2a'_lS_l+2a_l(\sum_{k=1}^l S_k)',\0\\
&&\dt S_l=2a'_1+2S_lS'_l.\0
\b
We see that all the fields $S_l(1\leq l\leq N-1$)
have the same equations of motion, so we may simply identify them, and
denote them $S$. Then our dispersionless equations are
\a
&&\dt a_l=2a'_{l+1}+2a'_lS+2la_lS',\0\\
&&\dt S=2a'_1+2SS'.
\label{3.1gendis}
\b
We remark that the same conclusion can be reached in full generality
if we start from the matrix models \cite{BX2}.
This set of equations is our starting point. Hereafter we will
analyse their Hamiltonian structure and the associated algebras.

\subsection{Bi--Hamiltonian structure}

First we notice that eqs.(\ref{3.1gendis}) can be
reexpressed as Hamiltonian equations
\a
\dt f=\{f, H_3\}_1=\{f, H_2\}_2.
\b
where $f$ stands for $a_l$ and $S$, and
\a
H_2=\int (a_2+a_1S)dx,\qquad
H_3=\int(a_3+2a_2S+a_1S^2+a_1^2)dx.
\b
The first Poisson brackets are
\a
&&\{a_i(x), a_j(y)\}_1=[(i-1)a_{i+j-2}\d+(j-1)\d a_{i+j-2}]\delta(x-y),
\quad 2\leq i,j\leq N-1.\0\\
&&\{a_1(x), S(y)\}_1=\delta'(x-y).\label{dispb1}
\b
while all the other vanish.

The second Poisson brackets are
\a
\{S(x), S(y)\}_2&=&{{N}\over {N-1}}\d \delta(x-y),\quad\quad\quad
\{S(x), a_1(y)\}_2=\d S\delta(x-y)\0\\
\{S(x), a_j(y)\}_2&=&{{N-j}\over {N-1}}\d a_{j-1}\delta(x-y),
\quad 2\leq j\leq N-1;\0\\
\{a_1(x), a_j(y)\}_2&=& \Big( a_j(x) \d +j \d a_j\Big)\delta(x-y),\0\\
\{a_i(x), a_j(y)\}_2&=&\bigg[(ia_{i+j-1}\d+j\d a_{i+j-1})
-(i-1){{N-j}\over {N-1}}a_{i-1}\d a_{j-1}+\label{dispb2}\\
&&+\sum_{l=1}^{i-2}\Bigl((i-l-1)a_{i+j-l-2}\d a_l+(j-l-1)a_l
\d a_{i+j-l-2}\Bigl)+\0\\
&&+(i-1)a_{i+j-2}\d S+(j-1)S\d a_{i+j-2}
\bigg]\delta(x-y), \quad\quad i,j\geq 2.\0
\b
One can prove that these two Poisson brackets satisfy the Jacobi identities.
So we have found a new integrable system. We remark that with respect to
these two Poisson brackets, each subset of fields
$\{S, a_1, a_2,\ldots,a_l\},~(l\leq N-1)$ form a subalgebra.

Once again, as was done several times before, this new system can
be given a most useful Lax pair representation.
We introduce the `operator'
\[
\ell=p+\sum_{l=1}^{N-1} {{a_l}\over{(p-S)^l}}.\label{dislaxgen}
\]
and use the second prescription (\ref{presc2}) above. Then
eqs.(\ref{3.1gendis}) can be written as
\[
\dt \ell=\{\ell^2_+, \ell\}.
\]
and, in general, we can introduce an infinite series of flows
\[
\ddtr \ell=\{\ell^r_+, \ell\},\qquad 1\leq r.
\]

\subsection{The Dirac reduction}

The only possible reduction is in this case
\a
S=0,\0
\b
The improved second Poisson brackets are
\a
\{a_i(x), a_j(y)\}_{2 D}\,&=&\,
\bigg[ia_{i+j-1}\d+j\d a_{i+j-1} -i{{N-j}\over{N}}a_{i-1}\d a_{j-1}
+\label{diswn'}\\
&&+\sum_{l=1}^{i-2}\Bigl((i-l-1)
a_{i+j-l-2}\d a_l+(j-l-1)a_l\d a_{i+j-l-2}\Bigl)\bigg]\delta(x-y).\0
\b
The reduced equations of motion are
\a
\frac{\partial a_i} {\partial t}= 2a'_{i+1}- 2
\frac {N-i}{N}a_{i-1}a'_1  \label{emkdvn}
\b

The above coincide with the Poisson brackets and equations of motion
of the dispersionless $N$--th KdV hierarchy. To see this, let us consider
the $N$--th KdV Lax operator in the dispersionless limit
\def\adis{{\cal A}}
\a
\adis=p^{N+1}+\sum_{l=1}^N w_l p^{N-l}
\b
The dispersionless equations of motion are
\a
\ddtr \adis=\{ \adis^{{r\over{N+1}}}_+, \adis\}.\label{diskdv}
\b

The Poisson structures become
\[
\{w_i(x), w_j(y)\}_1=
\Bigl[(i-N)w_{i+j-N-1}\d+(j-N)\d w_{i+j-N-1}
-N\delta_{i+j,N}\d\Bigl]\delta(x-y)
\]
and
\a
\{w_i(x), w_j(y)\}_2&=&
\bigg[iw_{i+j-1}\d+j\d w_{i+j-1}-i{{N-j}\over{N}}w_{i-1}\d w_{j-1}
\label{diswn}\\
&&+\sum_{l=1}^{i-2}\Bigl((i-l-1)
w_{i+j-l-2}\d w_l+(j-l-1)w_l\d w_{i+j-l-2}\Bigl)
\bigg]\delta(x-y).\0
\b

If we identify
\[
a_i=w_i,\quad\quad\quad i\geq 1
\]
we see that the last bracket coincides with (\ref{diswn'}), which
therefore constitutes a realization of the $w_N$ algebra. The equations
of motion trivially coincide.

In this way we have got another example of a reduction that ends up
with an $N$--th \kdvh, as expected on the basis of the general statements
of the previous section.
We finally remark that, if we start from the dispersionless limit of
${\cal S}_N^k$, with $0\leq k< N-1$, and make the reduction $S=0$
we end up always with the same system, i.e. the dispersionless $N$--th
KdV hierarchy.

{\bf Acknowledgements.} One of us (C.--S. X.) would like to thank Q.P.Liu
and Z.Y.Zhu for discussions and ICTP and SISSA for hospitality.

\end{document}